# Transfer Learning from Monolingual ASR to Transcription-free Cross-lingual Voice Conversion

*Che-Jui Chang*[1]

[1]Red Pill Lab
cjerry1243@gmail.com

## Abstract

Cross-lingual voice conversion (VC) is a task that aims to synthesize target voices with the same content while source and target speakers speak in different languages. Its challenge lies in the fact that the source and target data are naturally non-parallel, and it is even difficult to bridge the gaps between languages with no transcriptions provided. In this paper, we focus on knowledge transfer from monolingual ASR to cross-lingual VC, in order to address the content mismatch problem. To achieve this, we first train a monolingual acoustic model for the source language, use it to extract phonetic features for all the speech in the VC dataset, and then train a Seq2Seq conversion model to predict the mel-spectrograms. We successfully address cross-lingual VC without any transcription or language-specific knowledge for foreign speech. We experiment this on Voice Conversion Challenge 2020 datasets and show that our speaker-dependent conversion model outperforms the zero-shot baseline, achieving MOS of 3.83 and 3.54 in speech quality and speaker similarity for cross-lingual conversion. When compared to Cascade ASR-TTS method, our proposed one significantly reduces the MOS drop between intra- and cross-lingual conversion.

## 1. Introduction

Voice conversion (VC) is a task that converts a source speech to a target speech while the content remains unchanged. Recent approaches, including style transfer based (ST-based) [1, 2, 3], VAE-based [4, 5, 6, 7, 8], and Seq2Seq based [12, 14, 16, 17] methods, have been proved powerful in both parallel and non-parallel datasets for intra-lingual conversion. However, the cross-lingual conversion task remains a challenging topic. In such a setting, source and target speech are from different languages, which creates a natural gap while doing the conversion mapping. Most approaches that are proved effective in intra-lingual VC tasks would suffer from this gap problem. For instance, ST-based methods learn the direct mappings from source to target and from target to source using cycle-consistent loss, but the assumption of consistency could be invalid because the target language is different from the source. Specifically speaking, an acoustic pronunciation in one language may not exist in another. Besides, VAE-based methods are usually built to learn disentangled content and speaker features through an encoder-decoder network. During training, the content and speaker features are extracted from the same speech while in testing, the content is from source and the speaker information is from target. This mismatch would confuse the decoder as it only learns to reconstruct speech from the content features in its own language. Though the mismatch can be partly addressed by training an adversarial network that ensures the content feature is both speaker-independent and language-independent, it requires longer training time and still remains unjustified.

To address cross-lingual VC tasks, we need to be aware that every language has its own mapping between content units and acoustic pronunciations. When we train a model that does the mapping for one language (usually referred to as target language) and try to feed into the model the content from another (source language), this mismatch in content units often leads to poor audio quality and strange pronunciations in the converted speech. On the other hand, if the content representations of the source language (ex. phonemes or characters) are used deterministically for modeling target (foreign) speech, the model itself could not learn the mapping effectively. There will also exist a gap between the content inputs used for training and testing. Therefore, we believe the difference in representations of content units of each language is the major obstacle of resolving cross-lingual VC.

In this paper, we aim to deal with the content mismatch problem by a speaker-independent (SI) acoustic model. This model is trained on larger ASR datasets in the source language and then transferred to our cross-lingual VC system. The goal of this SI acoustic model is to extract non-deterministic and shared content representations between source and target speech. Here we call them phonetic features as they are strongly indicative of phonemes. We assume the process of learning phonetic features in the earlier ASR task eliminates speaker information and retains sufficient acoustic content information. During training, the phonetic features serve as content input to the conversion model. In testing, the same SI model is used, which alleviates the content mismatch between the source and target.

Our approach to VC tasks is to find a common SI representation between the source and target and then use it as content input to synthesize speech. To this sense, VC is analog to text-to-speech (TTS) conversion because both of their goals are modeling the mapping of contents and speech. Inspired by this idea, we are able to take advantage of state-of-the-art TTS architectures. For example, Tacotron [9] is a Seq2Seq model with attention that predicts the mel-spectrogram from a character or phoneme sequence. Also, transformer-based TTS models [11] have been proposed to generate high-quality speech while speeding up the training process. In this paper, we proposed our VC system based on Tacotron 2 [10] with a few modifications in the input sequence, attention module, and stop token prediction. We also build a many-to-many conversion system based on the multi-speaker TTS method [13].

The contribution of this paper is as follows:

- We transfer knowledge from monolingual ASR to address the content mismatch problem for cross-lingual VC.

- Our method is transcription-free because we do not need transcriptions for the speech in VC datasets.
- Our method outperforms cascade ASR-TTS and Zero-shot methods and reaches MOS of 3.83 and 3.54 in quality and similarity respectively, with only 0.45 and 0.57 lower than intra-lingual conversion setting.
- We open-source our implementation[1]. Audio samples can be found here[2].

## 2. Related Work

### 2.1. Non-parallel VC

Non-parallel VC tasks often rely on a content extractor to provide speaker-independent features as input to the conversion model. The VAE-based methods [4, 6, 7] aim to train a content encoder and an acoustic decoder together. To ensure the content encoder learns only the content information, some studies [6, 8] used feature disentanglement method with adversarial training. Additional adversarial network could also be applied at the output of the decoder [4], which is beneficial in generating speech with higher quality.

Moreover, content extractors could also be trained separately from the conversion itself. These methods often rely on off-the-shelf ASR toolkits to provide representations of speech units as content features. Previous works [5, 15, 16] have successfully applied phonetic posteriorgrams, PPGs, as input to the conversion model. PPGs are better than recognized word or phoneme sequences since they contain more information of source speech, including pronunciation and prosody. Using PPGs as input, any Seq2Seq modeling technique can be directly applied. For instance, these papers [17, 12] are based on recurrent neural networks with attention mechanisms and transformers respectively.

### 2.2. Cross-lingual conversion

Cross-lingual VC could be viewed as a transfer in voice styles. StarGAN-VC [1] and StarGAN-VC2 [2] are proposed to use cycle-consistent GANs for spectrum conversion between the source and target. The conversion model has to learn jointly the conversion of acoustic characteristics and the preservation of linguistic contents, which would be difficult for cross-lingual VC. The authors of Res-StarGan-VC [3] proposed a novel method that simplifies the learning objectives of the generator. The conversion model only needs to learn the change of acoustic characteristics between source and target. It also helps to speed up the training process.

Previous works [14, 18] have attempted to address cross-lingual VC using a separately-trained content feature extractor and speaker encoder. [18] uses bidirectional LSTMs to convert the bilingual PPGs (bPPGs) of source speech to acoustic features. [14] further extends the method by a jointly-learned speaker encoder network. Though bPPGs have been proved better than monolingual PPGs (mPPGs) in this paper [18], we show in our paper that mPPGs and latent phonetic features are still good options. This is probably because the capacity of our conversion model is much higher. Furthermore, the bPPG method requires ASR training for every language. This is impractical for those languages where resources are limited or difficult to access.

## 3. Method

Our proposed system has the following components: monolingual acoustic model, speaker encoder, Seq2Seq based conversion model, and neural vocoder, as shown in Figure 1. First, we pass the input speech to the monolingual acoustic model. The extracted phonetic features are then concatenated with prosody features, log energy and zero crossing rate, to serve as input to the conversion model. Our conversion model takes as input the features as well as the speaker embeddings extracted from the speaker encoder and outputs mel-spectrograms. Note that during the training phase, the phonetic features and speaker embeddings are extracted from the same speaker, while in the testing phase, the former are from source speech and the latter are from target speech. Last, we apply a neural vocoder to generate waveforms from the predicted mel-spectrograms.

### 3.1. Knowledge Transfer

In this paper, we use phonetic features as the shared content representations between the source and target. This is achieved by training a monolingual acoustic model on a larger dataset and then feeding all the speech into the same content extractor, regardless of their language. Though the monolingual phonetic features may not be perfect to represent the content of foreign speech, our later experiments find them good enough to contain sufficient content information and bridge the language gap. The idea that we apply the monolingual acoustic model to extract content information for foreign languages can be interpreted as knowledge transfer from monolingual speech recognition to voice conversion for foreign speakers. This is particularly useful when any data or resource of foreign speech is insufficient.

Moreover, our proposed VC system can be easily scaled up to multi-speaker one by providing an additional speaker embedding to the system. To do this, we transfer knowledge from speaker verification tasks [13]. Our multi-speaker conversion model relies on additional speaker information input to specify whose speech we want to synthesize. We first trained a speaker encoder on a larger dataset and then used it to extract a fixed-length embedding as speaker input to the conversion model.

#### 3.1.1. Monolingual SI acoustic model

We train our own acoustic model for phoneme recognition. The model consists of 3 bi-directional LSTM hidden layers and a fully-connected output layer. The dimension of LSTM all layers is set as 512. The input mfcc feature is concatenated with neighboring 7 features in both the left and right context, forming a 600-dimension input vector. For simplicity, we adopted 70 context-independent phoneme units as labels. We train our model on LibriSpeech [21] datasets, which is preprocessed by HTK forced alignment toolkit [23] to convert transcriptions to frame-aligned labels.

The training stops as it reaches 8.9% in phoneme error rate in dev-clean dataset. We experiment two phonetic features: one is the output of the acoustic model, namely the mPPG, whereas the other is the output of the last LSTM layer, namely, the deep phonetic feature (DPF).

---

[1] https://github.com/cjerry1243/TransferLearning-CLVC

[2] https://drive.google.com/drive/folders/1q9ZF8BatBItM9IwZ-YEIUG2Dde-Cz7Ha?usp=sharing

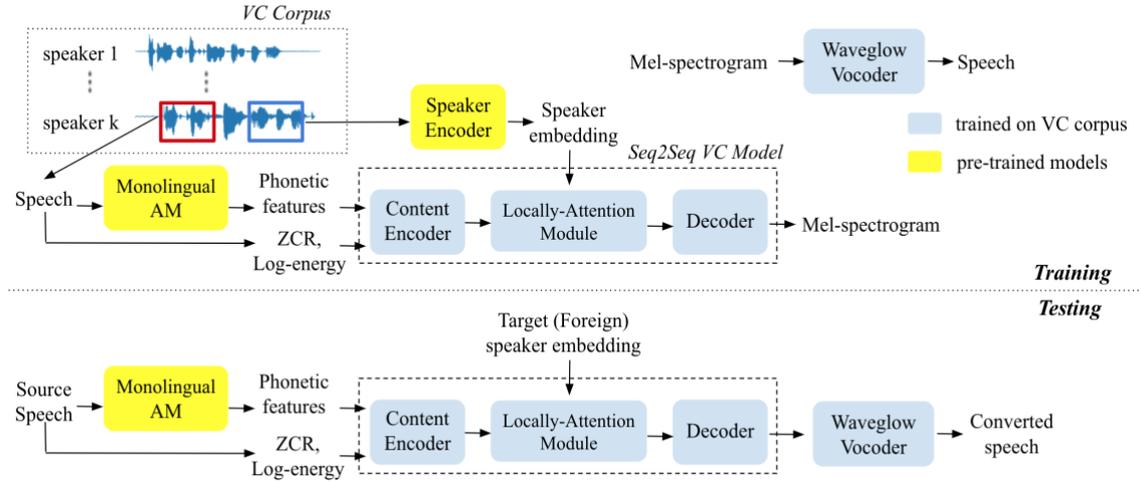

Figure 1: *Our proposed VC system*

### 3.1.2. Speaker encoder

Our speaker encoder is the same as the d-vector [19] system. The model is trained on VoxCeleb2 [22] dataset. The dimension of the embedding space is 256. As the training stops, our model reaches 7.4% in equal error rate in the test set.

### 3.2. Conversion model

Our conversion model is based on Tacotron 2 [10], a sequence-to-sequence model with an attention mechanism. The content encoder takes as input the phonetic and prosody features. The output features of the encoder are concatenated with speaker embeddings for all time steps. The attention module then learns a weighted combination of the concatenated features for decoding. Lastly, the decoder predicts mel-spectrograms.

Since the Tacotron 2 model is originally designed for text-to-speech synthesis, the length of its encoder input is usually shorter. If we replace the short text sequence with the long phonetic features as input without any modification, the model could not learn the attention mapping well. Therefore, we apply a locally-attention module with limited receptive fields of the encoder features. In fact, the pronunciation of a speech unit only depends on its local neighbors. In our experiments, we set the window size as 30 in both the left and right context. We also found this speeds up our training process.

Besides, we eliminate the stop token prediction module of Tacotron 2 because the output length of our conversion model is the same as the input length. During the testing phase, we stop the decoder prediction as the output length is equal to the length of source speech.

### 3.3. Neural Vocoder

We adopt Waveglow [20] as our vocoder. It's a flow-based generative model that learns the transformation from Gaussian normal distribution to speech data. With invertible operations and tractable determinants, Waveglow can be trained directly by maximizing the log-likelihood of the observed raw speech. It also accepts mel-spectrogram as conditional input, allowing controllable speech synthesis.

We first pretrained multi-speaker Waveglow on LibriTTS [25] dataset and then finetuned the vocoder on all provided VC training data. The window length and hop length are 32 ms and 10 ms respectively. The dimension of mel-spectrogram is 80. The vocoder is trained on ground-truth spectrograms, while in testing, the predicted ones are passed to the model to reconstruct waveform of target speech.

## 4. Experiments

We experiment our method on Voice Conversion Challenge 2020 (VCC 2020) [28] datasets, where 8 English, 2 Finnish, 2 German and 2 Mandarin speakers are provided. Each has 70 utterances, recorded in 24kHz. In the challenge, 4 English speakers are designated as source speakers, while others are target. For the intra-lingual conversion task (task 1), there are 16 source-target pairs, whereas for the cross-lingual conversion task (task 2), there are 24 pairs. For simplicity, we trained one conversion model and vocoder for all the pairs, except we compare ours with other speaker-dependent (SD) baselines.

In our experiment, we compare our proposed method with two baselines. We ask our reviewers to rate the speech quality and speaker similarity of the converted audio using mean opinion score (MOS) ranging from 1 to 5, in increments of 0.5. These are the baseline systems:

- **Cascade ASR-TTS** [26, 27] is a concatenation of transformer based ASR and TTS models. The ASR model is also trained on LibriSpeech dataset and the TTS model is first pretrained on LibriTTS dataset and then finetuned on VCC 2020 dataset. Note that this conversion model is a speaker-dependent one.
- **Zero-shot VC** is the same multi-speaker conversion model but trained on VCTK [24] dataset. We use the speaker encoder to extract speaker embeddings from target speakers in VCC 2020 dataset and then feed the embeddings to the conversion model.

We eliminate the silence in the head and tail of every utterance for audio preprocessing. We randomly extract several 10 seconds of audio segments for every speaker and obtain their speaker embeddings by our speaker encoder. During training, we randomly select one of the stored embeddings as conditional input to the conversion model.

## 5. Results

We present the MOS result of different methods in Table 1. For multi-speaker conversion, our proposed method achieves slightly lower speech quality but better similarity than the zero-shot baseline. This is because, compared to our method, the zero-shot one is trained on a larger English dataset, and the extracted speaker embedding from foreign speech does not decrease the speech quality. However, it fails to perform good adaptation to unseen or out-of-domain speakers and leads to a lower similarity score. This phenomenon is also observed in multi-speaker TTS [13] models. Besides, our two different content features, mPPG and DPF, obtain similar results. The DPF method is slightly better than mPPG. One possible reason is that mPPG only contains linguistic information, ex. probability distribution of phonemes, while on the other hand, the high-dimensional DPF preserves some prosody information that might be helpful for the conversion model.

Table 1: *MOS of different methods for intra- and cross-lingual conversion pairs.*

| Method | Quality | Similarity |
| --- | --- | --- |
| Zero-shot VC | 3.50 (0.41) | 2.57 (0.75) |
| Ours (mPPG) | 3.21 (0.47) | 2.91 (0.76) |
| Ours (DPF) | 3.37 (0.42) | 2.97 (0.72) |
| Cas-ASR-TTS (SD) | 2.15 (1.07) | 2.95 (1.02) |
| Ours (DPF-SD) | **3.97** (0.52) | **3.70** (0.70) |
| GT | 4.64 (0.23) | 4.69 (0.30) |

We further exploit DPF to train a speaker-dependent conversion model for each target speaker. Our method achieves an average of 3.97 in quality and 3.70 in similarity for task 1 and task 2 combined. Cascade ASR-TTS obtains 2.15 and 2.95. The source speech (ground truth) is rated the highest, with MOS of 4.64 and 4.69.

We observed that the scores of Cascade ASR-TTS are much different between intra- and cross-lingual conversion. This is also the main cause of their high standard deviation, as shown in Table 1. Table 2 further shows the MOS of intra- and cross-lingual conversion for Cascade ASR-TTS and ours. Cascade ASR-TTS obtains 3.79 and 4.16 for intra-lingual VC. However, it fails to do cross-lingual tasks well, obtaining 1.66 and 2.59 in MOS. We further found most of its cross-lingually converted speech is unintelligible. According to the objective evaluation in VCC 2020 [28], the WER of Cascade ASR-TTS method in intra-lingual conversion is less than 10%, but its WER in cross-lingual conversion increases up to more than 30%. On the contrary, our proposed method (DPF-SD) achieves 3.83 and 3.54 in MOS, with just a small decrease compared to intra-lingual VC.

Table 2: *MOS of intra- and cross-lingual conversion.*

| Method | Quality (intra/cross) | Similarity (intra/cross) |
| --- | --- | --- |
| Cas-ASR-TTS (SD) | 3.79/1.66 | 4.16/2.59 |
| Ours (DPF-SD) | 4.28/**3.83** | 4.11/**3.54** |

The major difference between Cascade ASR-TTS method and ours is the usage of different content inputs. Cascade ASR-TTS first transcribes speech into texts and uses texts as input to its conversion model. Its ASR model is specifically trained and optimized on English corpus, so for intra-lingual (English-to-English) conversion, the transcribed texts are very accurate for both source and target speech. The content input remains accurate and consistent during the training and testing phase. However, when we use it as the content extractor for foreign speech, the content input, first, has some error. Second, as we optimize our conversion model based on the erroneous texts, the inaccuracy is amplified. Moreover, in testing stage, we do not pass to the model the same texts used for training, but the accurate texts obtained from source speech. The amplified transcription error together with the content mismatch problem deteriorates the quality of converted speech.

Table 3: *MOS of all language-to-language conversion pairs.*

| Language (Source-Target) | Quality | Similarity |
| --- | --- | --- |
| English-English | 4.28 (0.48) | 4.11 (0.60) |
| English-Finnish | 3.78 (0.42) | **3.68** (0.58) |
| English-German | **4.05** (0.42) | **3.69** (0.64) |
| English-Mandarin | 3.67 (0.52) | 3.27 (0.67) |

Our proposed method also transfers knowledge from English ASR tasks, but we use the intermediate output feature as content input. The phonetic features are, assumably, shared representations across languages. They contain sufficient content information while not optimized too much on English corpus. Table 3 shows the result of all language-to-language VC pairs by our DPF-SD method. For English-English conversion, our model has the best quality and similarity score. This proves that the intra-lingual knowledge transfer between ASR and VC is the most effective. The English-German conversion has the closest performance to English-English conversion. This is mainly because these two languages are similar in phonology. The scores of English-Mandarin conversion are the lowest among all. The differences between the two languages in phonology, tone, and talking speed makes it difficult to generate high-quality English speech for Mandarin speakers. We also observed that not only the conversion error of our model but the language difference affects similarity scores. An additional bilingual speech for each speaker should be included in the listening review to account for the similarity loss caused by the conversion process.

## 6. Conclusions

We propose a novel method for cross-lingual VC that uses pre-trained monolingual acoustic model as its content extractor. Our experiment suggests the shared content representations are effective in bridging content gaps for different languages and lead to higher quality and similarity scores compared with our baselines. Most importantly, our transcription-free approach could be easily applied to languages with limited labels or resources.

In the future, we will focus our research on self-supervised or unsupervised speech representation learning, in an effort to learn more powerful language-independent content representations that could be used for multi-lingual VC, speech synthesis for data-limited languages, or other few-shot learning applications.